\documentstyle[preprint,aps,epsfig]{revtex}

\def\x{{\bf x}}
\def\y{{\bf y}}

\def\be{\begin{equation}}
\def\ee{\end{equation}}
\def\bea{\begin{eqnarray}}
\def\eea{\end{eqnarray}}

\title{Total cross section for $p-d$ breakup below 30 MeV}
\author{A.~Kievsky$^{a,b}$,C.~R.~Brune$^{b,c}$,M.~Viviani$^a$}
\address{$^a$Istituto Nazionale di Fisica Nucleare, Via Buonarroti 2, 56100 Pisa, Italy}
\address{$^b$Triangle Universities Nuclear Laboratory, Duke University, Durham,
         NC 27708, USA}
\address{$^c$Department of Physics and Astronomy, University of North Carolina at
         Chapel Hill, Chapel Hill, NC 27599-3255, USA}

\begin{document}

\maketitle
\begin{abstract}
 The total cross section for p-d breakup is studied in terms
 of the elastic ${\cal S}$--matrix through the unitary condition. 
 Calculations using the complex Kohn variational method along with the
 Pair Correlated Hyperspherical Harmonic basis are presented. 
 The results have been restricted to energies below $E_p=30$~MeV where Coulomb
 effects are expected to be sizable and are compared to the existing data. 
 Two different measurements have been found in the literature:
 40 years ago, Gibbons and Macklin (1959); and 26 years ago, Carlson 
 {\sl et al.} (1973). The calculations are found to be
 in reasonable agreement with these old data, though a discrepancy
 is observed near the deuteron breakup threshold. Moreover, a detailed
 analysis of the contributions to the observable from different
 partial waves has been performed.
 Unexpectedly, the main contribution for a wide
 range of energies has been detected in the $J=3/2^-$ state. 
 
\end{abstract}

 keywords: N-d scattering; breakup cross section     \\
 corresponding author: Alejandro Kievsky     \\
Istituto Nazionale di Fisica Nucleare, Via Buonarroti 2, 56100 Pisa, Italy     \\
tel. +39-050-844546, FaX  +39-050-844538, e-mail: kievsky@pisa.infn.it

\newpage

 Studies in the three-nucleon (3N) continuum are based
 mainly in N-d scattering experiments.
 The deuteron
 is the only existing bound state in the two-nucleon (2N)
 system with a binding energy $B_d=2.225$ MeV. 
 Therefore, for incident nucleon energies below $3.337$ MeV,
 or incident deuteron energies below $6.675$ MeV, there are
 no open channels (disregarding the very low probability for
 radiative capture or bremsstrahlung)
 and the reaction goes through the elastic channel.
 Accordingly the elastic scattering matrix, the ${\cal S}$--matrix, is
 unitary. 

 For energies above the deuteron breakup threshold (DBT) three free
 nucleons are present in the outgoing channel. For example, in p-d
 scattering a free neutron can be observed only above the
 DBT (neutron production)
 whereas for the n-d reaction the observation of a free proton is only
 possible above the DBT.
 The elastic ${\cal S}$--matrix is no longer unitary
 and the missing flux in the elastic channel is related to the total
 breakup cross section.

 The total breakup cross section accounts for all possible configurations
 of the three free outgoing particles and its expression is given for
 example in Ref.~\cite{LT} in terms of the ${\cal T}$-matrix
 elements connecting the elastic channels to the inelastic ones:
 \begin{equation}
      \sigma_b(N-d)= {\pi\over k^2}{1\over (2I_1+1)(2I_2+1)}
      \sum_J (2J+1)\sum_{\alpha\beta} |T^J_{\alpha\beta}|^2  \ .
 \label{eq:sigmab}
 \end{equation}
 The index $\alpha$ labels the elastic channels while $\beta$
 runs over all possible configurations of the three outgoing
 nucleons. In addition
 $k^2={2\mu \over \hbar^2}E_0$, where $\mu$ is the reduced 
 nucleon-deuteron mass, $E_0$ is the center of mass energy, and $I_1$ ($I_2$)
 is the spin of the incident particle (target). For each state with
 total angular momentum $J$ the flux conservation in channel
 $\alpha$ imposes the unitary condition

 \begin{equation}
       \sum_{\alpha'}|{\cal S}^J_{\alpha\alpha'}|^2 +
       \sum_\beta |T^J_{\alpha,\beta}|^2 = 1  \ .
 \label{eq:sigmau}
 \end{equation}

 Experimental measurements of the total breakup
 cross section $\sigma_b$ are scarce
 and have been performed many years ago with limited accuracy.
 For the n-d system an indirect determination
 is possible through the subtraction of the total elastic cross
 section $\sigma_{el}(n-d)$ from the total nuclear cross section
 $\sigma_{tot}(n-d)$.
 Existing $\sigma_{el}(n-d)$ data lack the desired accuracy producing
 a large uncertainty in $\sigma_b(n-d)$, particularly at
 low energies~\cite{schwarz}.
 Direct measurements of $\sigma_b(n-d)$ have also been done more
 than 20 years ago~\cite{sigmab-n}. The interest at that time was to produce
 data for comparison with the first theoretical attempts to solve
 the 3N continuum using semi-realistic N-N potentials.
 These early potential models
 gave a qualitative description of the data but
 the uncertainties of $\sim 10\%$
 were too large to make definitive conclusions.
 The present status of this observable is briefly given in Ref.~\cite{report} 
 (see page 184) where modern theoretical predictions are compared to the data. 
 The calculations in Ref.~\cite{report} have been performed
 by solving the Faddeev equations
 in momentum space using different realistic potential models. 
 The discrepancies observed at low energies have been attributed to the low
 quality of the data. In fact, since the total and elastic cross sections
 are well described by theory there is no reason to expect a discrepancy
 in the total breakup cross section.
 On the other hand, since the  the total, elastic, and breakup
 cross section are related by the unitary condition and the optical theorem,
 it is highly desirable to have a consistent description
 of all three observables. New determinations of $\sigma_b(n-d)$ could
 come from both direct or indirect measurements.

 The situation of the p-d total breakup cross section is somewhat different.
 In this case no indirect measurement is possible due to the Coulomb
 divergence in the elastic amplitude. Two direct measurements have been
 found in the literature. A very old one by Gibbons and Macklin
 (1959) below $E_p=5.6$ MeV~\cite{gibbons},
 with systematic uncertainties estimated to be $40\%$ at $E_p=3.5$~MeV
 and decreasing to $5\%$ at $E_p=5.5$ MeV.
 A second experiment was performed by 
 Carlson {\sl et al.} (1973) for proton energies between $22$ and 
 $46$~MeV~\cite{ivo}, with uncertainties between $8\%$ and $5\%$. 
 The experimental techniques utilized were quite different
 in the two measurements. 
 In the first $\sigma_b$ has been
 determined by measuring the total neutron yield
 while in the second one an attenuation technique has been used.
 These two measurements provide valuable information
 to be used in the comparison to a theoretical description of p-d scattering
 including the long-range Coulomb force. In particular 
 Coulomb effects are expected to be important in the Gibbons and Macklin
 experiment performed at energies near threshold.
 For example $\sigma_b(n-d)= (34\pm6)$~mb at $E_n=4.9$ MeV
 (Holmberg (1969)~\cite{sigmab-n}) to be compared to
 the measurement of Gibbons and Macklin, $\sigma_b(p-d)= 21.5\pm1.4$ at the
 same energy. The Coulomb force reduces the breakup cross section by 
 approximately $30\%$.
 At higher energies the effect is less evident: at
 nucleon energies of $22$ MeV the
 error bars of the n-d and p-d measurements start to overlap, and the
 effect seems to be extremely small above $40$ MeV.
 A collection of n-d and p-d data is given in Fig.~1.

 Recently the variational technique based on an expansion in terms of the
 Pair Correlated Hyperspherical Harmonic (PHH) basis has been extended to
 describe elastic N-d scattering above the DBT~\cite{KVR97,KRV99,VKR00}.
 The method consists of an expansion of the 3N wave function onto the PHH
 basis and the elastic elements of the ${\cal S}$--matrix are obtained 
 through the application of the Kohn variational principle in complex 
 form (see Ref.~\cite{kiev97} and references therein). 
 The inclusion of the Coulomb interaction in the context of
 the variational method using the PHH expansion gives no particular
 troubles provided the correct boundary conditions are imposed on the
 wave function in the asymptotic region. An extensive discussion 
 of the asymptotic form of the wave function in connection with the Kohn
 variational principle and the application of the PHH expansion
 is given in Ref.~\cite{VKR00}.
 In Ref.~\cite{KVR97} it was shown that the method
 reproduces the benchmarks of Ref.~\cite{bench} for n-d scattering whereas
 in Ref.~\cite{KRV99} the elastic cross section and polarization observables have been
 calculated for n-d and p-d scattering at nucleon energies of
 $5$ and $10$~MeV using
 realistic potentials. In the later case the applications to n-d
 scattering have been shown to be in close agreement to those given
 by the Bochum-Cracow group~\cite{report}.

 Applications to describe p-d
 scattering above the DBT are of particular interest due to the historical
 difficulties to manage the distortion introduced by the Coulomb force
 in the asymptotic region. In this context, observables sensitive to the
 long-range Coulomb interaction give an unique opportunity to test techniques
 devised to take into account such distortion. The p-d total breakup cross
 section is a well suited observable to make comparisons. The existing
 amount of data, though small, is sufficient to extract conclusion about
 the capability of the PHH technique to describe the charged 3N system
 above threshold. Moreover, as we shall see, it will be possible to evaluate
 the need for new and higher-precision measurements of $\sigma_b$.
 These measurements would
 increase our understanding of $\sigma_b$ and impose 
 stringent conditions on the theoretical methods and
 phase-shift analyses (PSA's).

 The breakup cross section $\sigma_b(N-d)$
 is given in Eq.~(\ref{eq:sigmab}) 
 in terms of the ${\cal T}$--matrix elements connecting
 the elastic and inelastic
 channels. Using the fact that those elements are related to the elastic
 ${\cal S}$--matrix elements through the unitary condition 
 of Eq.~(\ref{eq:sigmau}) it is possible
 to re-write the above equation in terms of the elastic matrix elements..
 In the case of N-d scattering the elastic ${\cal S}$--matrix corresponding
 to a state with total angular momentum $J$ is a $3\times 3$ matrix. Each
 matrix element ${}^J{\cal S}^{L'S'}_{LS}$ is labeled by the 
 sets of quantum numbers $[\alpha\equiv LS]$ coupled to $J$. Here $L$ is the 
 relative angular momentum between the incident 
 nucleon and the deuteron and $S$ is the total spin coming from the coupling
 of the spin of the deuteron $S_d=1$ to the spin $S_N=1/2$ of the nucleon.
 Accordingly, $S=1/2$ or $3/2$ and there are three possible couplings of
 $[LS]$ giving $J$ and conserving parity [the parity of the state is
 given by $(-1)^L$].
 In the case of $J=1/2^{\pm}$ there are
 two possible couplings and the ${\cal S}$--matrix is a $2\times 2$ matrix.
 Moreover for N-d scattering $(2I_1+1)(2I_2+1)=6$.
 From the above considerations Eq.~(\ref{eq:sigmab}) can be given in the
 more compact form:
 
 \begin{equation}
      \sigma_b(N-d)= {\pi\over k^2}{1\over 6}
               \sum_J (2J+1)tr\{I_J-{\cal S}_J{\cal S}_J^\dagger\}\ ,
 \label{eq:sigmac}
 \end{equation}
 where $I_J$ is the $3\times3$ identity matrix excepts for $J=1/2^{\pm}$
 which is the $2\times2$ identity matrix.
 Eq.(\ref{eq:sigmac}) gives $\sigma_b$ in terms of the elastic matrix
 elements at fixed $J$, the sum runs over all possible values of $J$
 and parity (the sum over different parities is understood). Although
 the sum runs from $J=0$ to infinity there is a rapid convergence due to
 the fact that each ${\cal S}_J$ matrix becomes closer to unitary as
 $J$ increase. High $J$ values correspond to high $L$ values and the
 centrifugal barrier prevents peripheral waves from participating actively
 in the breakup process. Moreover quartet states are also peripheral
 due to Pauli blocking. For each $J$ states there are two quartet states and
 one doublet state which in combination with the increasing $L$ values
 limit the number of terms in the sum to a tractable number 
 inside the energy range we are
 going to consider. In fact, below $30$ MeV states with 
 $J > 11/2^{\pm}$ make a negligible contribution to $\sigma_b$.
 Eq.~(\ref{eq:sigmac}) gives a very simple picture of the observable, as the 
 diagonal elements of the product matrix ${\cal S}_J{\cal S}_J^\dagger$ 
 are always in the range $(0,1)$. Therefore the quantity 
 $A_J=tr\{I_J-{\cal S}_J{\cal S}_J^\dagger\}$ divided
 by $tr\{I_J\}$ gives a measurement of the inelasticity in that state.
 The complete contribution  to the breakup cross section is obtained
 after taking into account the spin degeneracy $(2J+1)$. 
 
 In the present paper we present calculations of $\sigma_b(N-d)$ based
 on the PHH technique. Thorough details of the method are given in
 Refs.~\cite{KVR97,KRV99,VKR00} and only a brief discussion will be
 presented.
The starting point in the present calculations is the Kohn variational
principle in complex form, which states that the functional
\begin{equation}
  [{}^J{\cal S}^{L'S'}_{LS}]= {}^J{\cal S}^{L'S'}_{LS}+i
  \langle\Psi^-_{LSJ}|H-E|\Psi^+_{L'S'J}\rangle \ ,
  \label{eq:kohn}
\end{equation}
is stationary with respect to the trial parameters in
the three-nucleon scattering wave function. For an incident 
state with relative angular momentum $L$, spin $S$ and total angular momentum
$J$ it is
\begin{equation}
\Psi^+_{LSJ}=\sum_{i=1,3}\left[ \Psi_C(\x_i,\y_i)+\Omega^+_{LSJ}(\x_i,\y_i)
             \right] \ ,\label{eq:psip}
\end{equation}
and its complex conjugate is $\Psi^-_{LSJ}$.
The first term, $\Psi_C$, describes the
system when the three--nucleons are close to each other and, asymptotically,
an outgoing three-particle state.
Each amplitude $\Psi_C(\x_i,\y_i)$, where $\x_i,\y_i$ are
the Jacobi coordinates corresponding to the $i$-th permutation, has
total angular momentum $JJ_z$ and total isospin $TT_z$ and it is
decomposed into channels labelled by angular-spin-isospin quantum
numbers ($\beta$--channels). 
The remaining two-dimensional amplitude is expanded in terms of the PHH basis
\begin{equation}
     \phi_\beta(x_i,y_i) = \rho^{\ell_\beta+L_\beta-5/2}f_\beta(x_i) 
     \left[ \sum_K u^\beta_K(\rho)\, {}^{(2)}P^{\ell_\beta,L_\beta}_K(\phi_i)
     \right] \ ,
\label{eq:PHH}
\end{equation}
where the hyperspherical variables are defined by the relations
$x_i=\rho\cos{\phi}_i$ and $y_i=\rho\sin{\phi}_i$,
$f_\beta(x_i)$ is a pair correlation function and
${}^{(2)}P^{\ell,L}_K(\phi)$ is a hyperspherical polynomial.
The second term in the variational scattering w.f.  
describes the asymptotic  motion of a deuteron relative to the third
nucleon. It can be written 
in terms of the ingoing and outgoing solutions
of the asymptotic N-d Schroedinger equation.
\begin{equation}
\Omega^+_{LSJ}(\x_i,\y_i) =  \Omega^{in}_{LSJ}(\x_i,\y_i)-
 \sum_{L'S'}{}^J{\cal S}^{L'S'}_{LS}\Omega^{out}_{L'S'J}(\x_i,\y_i)  \ .
\end{equation}

For energies above the DBT the hyperradial functions $u^\beta_K(\rho)$
in Eq.(\ref{eq:PHH}) should describe an outgoing wave distorted by the 
Coulomb interaction between the two protons.
Accordingly, for $\rho\rightarrow\infty$, the hyperradial
functions behaves as 
\begin{equation}
  u^\beta_K(\rho) \rightarrow  \sum_{\beta'K'}
  (e^{-i\chi\log2Q\rho})^{KK'}_{\beta\beta'}B^{\beta'}_{K'} e^{iQ\rho}
    \label{eq:bc}
\end{equation}
where $Q^2=ME/\hbar^2$, with $M$ the nucleon mass, $E$ is the total
energy ($E=E_0-B_d$), and the $\chi$--matrix originates from 
the Coulomb potential. Extending the index $\beta$ to include the
hyperspherical quantum number $K$, the numerical constants
$B^\beta_K$ are the ${\cal T}$-matrix elements $T^J_{\alpha,\beta}$
defined in Eq.(\ref{eq:sigmab}).

 We have studied the convergence of the quantity $A_J$
 in terms of the
 partial wave channels used in the decomposition of the wave function for
 each $J^\pm$ state. Examples of the convergence can be found in 
 Ref.~\cite{VKR00}. The accuracy of the results given here is estimated
 to be better than $1\%$.
 The calculations have been done with the Argonne $v_{18}$ (AV18)
 interaction~\cite{av18} and in some cases the three-nucleon force (3NF) 
 of Urbana (UR)~\cite{urbana} has been included
 for the study of the sensitivity to 3NF's.

 In Fig.~2 the PHH results for $\sigma_b(p-d)$ 
 are shown together with the experimental p-d data.
 The theoretical calculations (open squares) are in good agreement with the
 data. There is a window between $6$ MeV and $22$ MeV with no data,
 just in the region where the observable is varying rapidly and reaches its
 maximum. Above $20$ MeV the calculations have been done at $22.7$ MeV
 and $28$ MeV, and are in close agreement with the measurements from
 Carlson {\sl et al.}~\cite{ivo}
 In Fig.~3 a detailed plot with semilogarithmic scale is given at the
 very low energies explored by Gibbons and Macklin~\cite{gibbons}.
 The theoretical results with the AV18 interaction (open triangles) 
 and AV18+UR interaction (open squares)
 are in agreement at the higher energies but start to deviate as
 the energy approaches
 the DBT. This region is of particular interest since Coulomb effects are huge
 and the breakup amplitude is attenuated by the Coulomb penetration factor
 $\exp(-2\pi\eta)$, with $\eta\propto 1/\sqrt{E_0-B_d}$.
 At $5$ MeV the PHH n-d results are also given in Fig.~3 for the 
 sake of comparison (diamond).
 The n-d data from Holmberg closest to $5$ MeV are 
 plotted also (open circles). The effect of the Coulomb interaction is
 clearly evident.  Moreover the calculations at $5$ MeV
 based on the variational method reproduce both
 the n-d and p-d breakup reactions. 

 Differences between theory and experiment are observed around $4$ MeV and
 below. These discrepancies could originate
 from an incomplete treatment of the
 electromagnetic interaction. In fact, the calculations included the Coulomb
 interaction but not other small electromagnetic parts of the interaction
 such as the vacuum polarization or the magnetic moment
 potentials. On the other hand
 systematic errors in the measurements are not unlikely --
 a new set of measurements will be highly desirable. 

 In Fig.~3 the effects of the 3NF's are also studied (open squares).  
 At these very low energies the breakup cross section is
 dominated by doublet states with $L=0,1$ since the probability
 to find the three nucleons close to each other is not reduced by either
 Pauli blocking or a high centrifugal barrier. The doublet ${}^2S_{1/2}$
 state is sensitive to 3NF's. For example calculations with AV18 or
 AV18+UR at $3$ MeV are $20\%$ different for the ${}^2S_{1/2}$ 
 phase-shift~\cite{KRTV96}.
 In this case we are referring to energies below the DBT
 where the phase-shifts are real but we can expect sensitivity to 3NF's
 also in the
 imaginary parts above the DBT. The results given in Fig.~3 show a 
 reduction of $\sigma_b$ of the order or $3\%$--$4\%$ when the AV18+UR
 interaction is used. The inclusion of the 3NF increases the binding
 energy of the three-nucleon system, and accordingly the inelasticity
 of the $J=1/2^+$ state is reduced, diminishing the
 total breakup cross section.

 Finally we want to discuss the contribution of the different states to 
 the total breakup cross section. 
 Each state contributes to $\sigma_b$ an amount equal to $(2J+1)A_J$.
 In Fig.~4 we plot this quantity up to $J=9/2^\pm$. It is interesting to
 note that above $6$ MeV the state $J=3/2^-$ gives by far the main
 contribution to the observable. This state has the optimum combination
 of a large $A_J$ value coming mainly from ${}^2P_{3/2}$ and the spin
 degeneracy $2J+1=4$. This is twice the value for $J=1/2$ and in fact
 the $J=3/2^-$ contribution is double the $J=1/2^-$ one. Although the
 $J=1/2^+$ state has the highest $A_J$ value the degeneracy is too
 low and its contribution is the largest one only below $6$ MeV. Above
 $24$ MeV other states start to have important contributions, showing the
 weakness of the centrifugal barrier as the energy increases. 
 In Fig.~5 the contributions to $\sigma_b$ are given for energies below
 $7$ MeV. The contributions calculated using the AV18+UR potential
 are also given (dotted line).
 As it was expected the $J=1/2^+$ contribution is principal
 below $6$ MeV and, more important, is the only one below $4$ MeV.

 The reduced numbers of parameters in description of $\sigma_b$ at low
 energy is of
 particular interest. At $4$ and $5$ MeV a complete set of p-d vector and
 tensor analyzing powers as well as p-d differential cross section have been
 measured~\cite{sagara,lynn}. These high-precision data can be used
 to perform a single-energy PSA in which some of the phases must be
 allowed to be complex. From the above analysis it is clear that only
 few phases or mixing parameters have sizable imaginary parts. In addition, the
 inclusion of $\sigma_b$ in the data base will impose further restrictions
 on the imaginary parts. 

 PSA is extremely useful for performing comparisons between theory
 and experiment. Discrepancies at the level of observables can be
 traced back to single phases or mixing parameters which in turn
 are directly related to specific parts of the interaction. Above the
 DBT phases and mixing parameters are complex doubling the number
 of variables in the search procedure. Old PSA's encountered difficulties
 finding a precise determination of the real and imaginary parts
 of the phases and mixing parameters~\cite{arvieux}. The high-precision
 p-d data obtained through the past years allowed for a better
 determination of the parameters through single energy PSA~\cite{KRTV96,lynn2}.
 These analyses were limited to energies below the DBT. The study of the
 breakup cross section in terms of partial waves will be helpful
 when PSA's are extended to higher energies.
 In this light it would be particularly useful to perform new experimental
 determinations of $\sigma_b(p-d)$ at energies between 6 and 22 MeV,
 where there are presently no experimental results.
 It would also be most advantageous to perform these measurements
 at the same energies where highly-accurate differential cross section
 and analyzing power data already exist~\cite{sagara}.

\begin{acknowledgements}
 
 One of the authors (A. K.) would like to thank the Triangle Universities 
 Nuclear Laboratory for hospitality and support where much of this work
 was performed.
 Moreover, the authors would like to thank many useful discussion
 with W. Tornow, E.~J.~Ludwig, H.~J.~Karwowski and S. Rosati.
\end{acknowledgements}

\newpage

FIGURE CAPTIONS

\noindent Fig.~1. Total breakup cross section for n-d (filled circles) and
 p-d (open circles) scattering up to $50$~MeV. Experimental data are from
 Ref.~\cite{schwarz,sigmab-n,gibbons,ivo}.

\noindent Fig.~2. Theoretical calculations of $\sigma_b(p-d)$ up to $30$ MeV
 (open squares). The experimental results of Gibbons and Macklin~\cite{gibbons}
 (open triangles) and Carlson {\sl et al.}~\cite{ivo} (open circles) are
 given for the sake of comparison.

\noindent Fig.~3. Total breakup cross section below 6 MeV. 
 Theoretical calculations
 for $\sigma_b$ are given at four energies using the AV18 potential 
 (p-d:open triangles, n-d:diamond) and AV18+UR potential (p-d:open squares). 
 The experimental data are from Gibbons and Macklin~\cite{gibbons}
 (p-d:filled circles) and Holmberg~\cite{sigmab-n} (n-d:open circles). 
 The solid line is a fit to the p-d data.

\noindent Fig.~4. The quantity $(2J+1)A_J$ for different values of 
 $J$ and parity
 between 5 and 28~MeV. Calculations have been done at the same 
 energies of Fig.~2. The solid lines are linear interpolations.

\noindent Fig.~5. Same as Fig.~4 below $7$ MeV. Calculations have been done
 using AV18 (solid line) and AV18+UR (dotted line).


\begin{thebibliography}{9}

\bibitem{LT} A.~M.~Lane and R.~G.~Thomas, Rev. Mod. Phys. 30 (1958) 257.

\bibitem{schwarz} P.~Schwarz {\sl et al.}, Nucl. Phys. A 398 (1983) 1;
        J.~D.~Seagrave {\sl et al.}, Ann. Phys. 74 (1972) 250.
 
\bibitem{sigmab-n} H.~C.~Catron {et al.}, Phys. Rev. 123 (1961) 218;
        M.~Holmberg, Nucl Phys. A 129 (1969) 327;
        G.~Pauletta and F.~D.~Brooks, Nucl. Phys. A 255 (1975) 267.
 
\bibitem{report} W.~Gl\"ockle {\it et al.}, Phys. Rep. 274 (1996) 107.
 
\bibitem{gibbons} J.~H.~Gibbons and R.~L.~Macklin, Phys. Rev. 114, (1959) 571.

\bibitem{ivo} R.~F.~Carlson {\sl et al.}, Lett. N. Cim. 8 (1973) 319.

\bibitem{KVR97} A.~Kievsky, M.~Viviani, and S.~Rosati,
        Phys. Rev. C 56 (1997) 2987.
 
\bibitem{KRV99} A.~Kievsky, S.~Rosati, and M.~Viviani,
        Phys. Rev. Lett. 82 (1999) 3759.
 
\bibitem{VKR00} M.~Viviani, A.~Kievsky, and S.~Rosati, 
        Few--Body Systems, in press.
 
\bibitem{kiev97} A.~Kievsky, Nucl. Phys. A 624 (1997) 125.

\bibitem{bench} J.~L.~Friar {\sl et al.}, Phys. Rev. C 51 (1995) 2356.

\bibitem{av18} R.B. Wiringa, V.G.J. Stoks and R. Schiavilla, Phys. Rev. C51 (1995) 38

\bibitem{urbana} B.S. Pudliner {\sl et al.}, Phys. Rev. C56 (1997) 1720

\bibitem{KRTV96} A.~Kievsky, S.~Rosati, W.~Tornow, and M.~Viviani,
        Nucl. Phys. A 607 (1996) 402.

\bibitem{sagara} K.~Sagara, {\sl et al.}, Phys. Rev. C 50 (1994) 576;
        S.~Shimizu, {\sl et al.}, Phys. Rev. C 52 (1995) 1193;
        K.~Sagara, private communication.

\bibitem{lynn} J.~Sowinski, D.~D.~Pun Casavant and L.~D.~Knutson,
        Nucl. Phys. A 464 (1987) 223.

\bibitem{arvieux} J.~Arvieux, Nucl. Phys. A 221 (1974) 253.

\bibitem{lynn2} L.~D.~Knutson, L.~O.~Lamm, and J.~E.~MacAninch,
        Phys. Rev. Lett. 71 (1993) 3762.

\end{thebibliography}
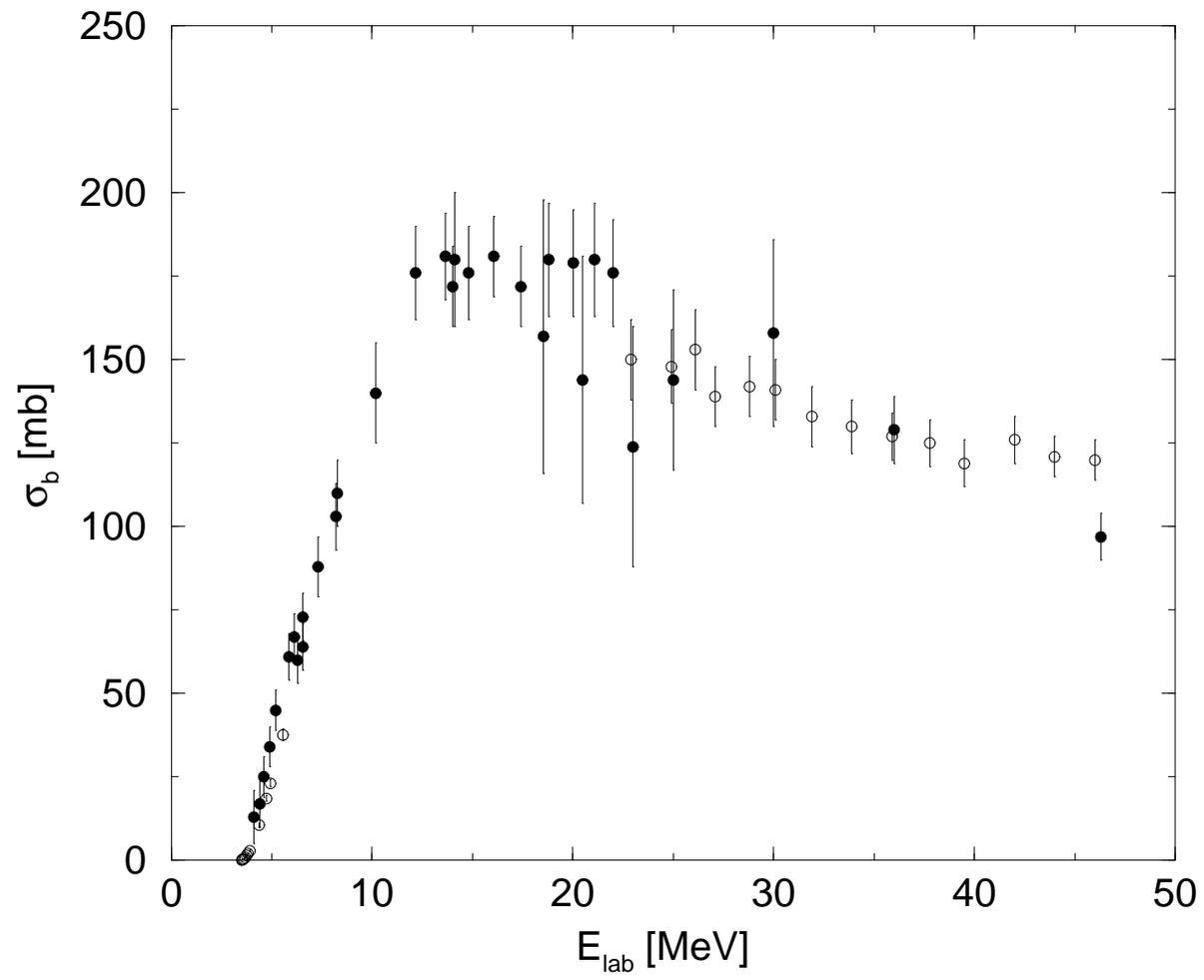
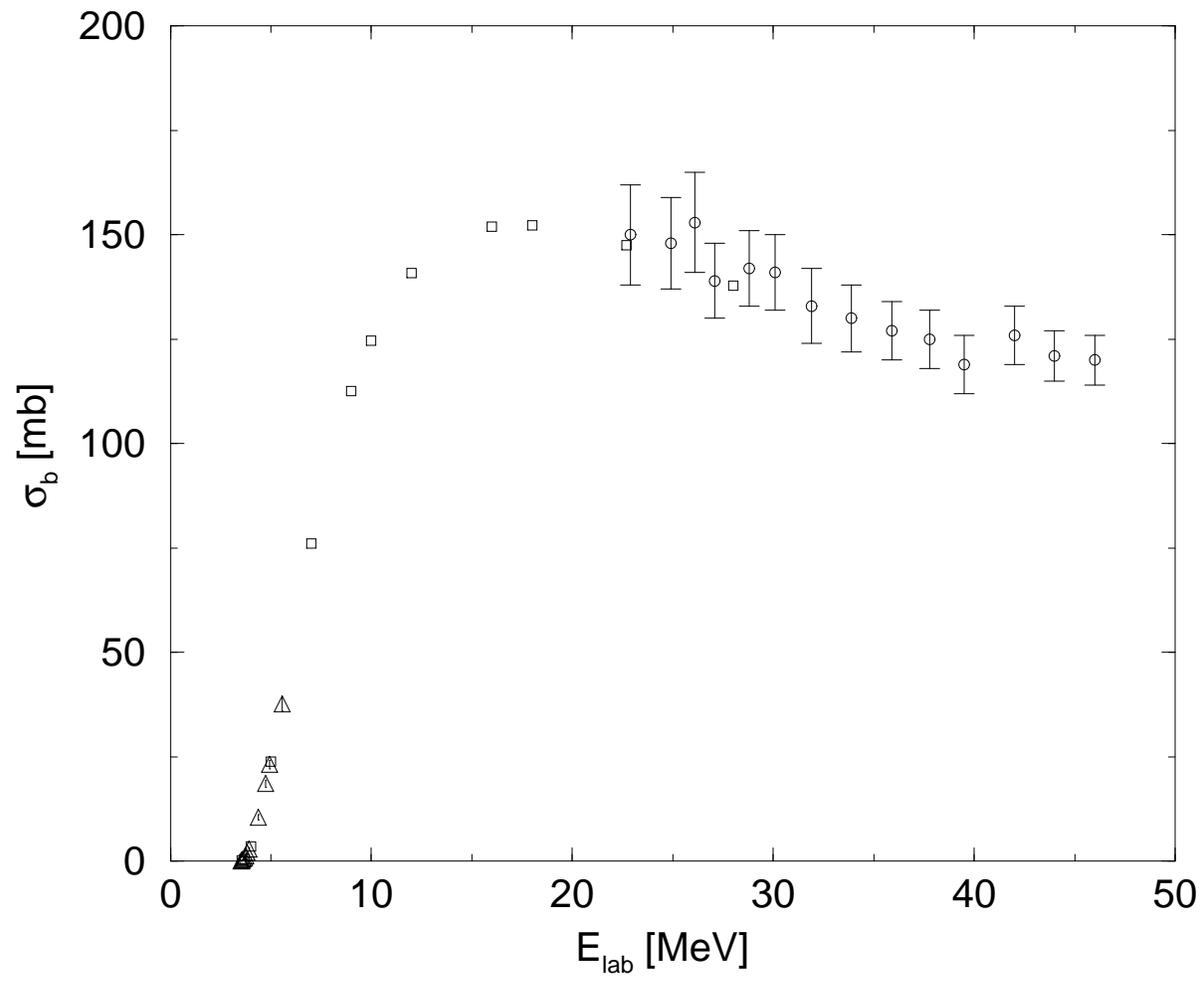
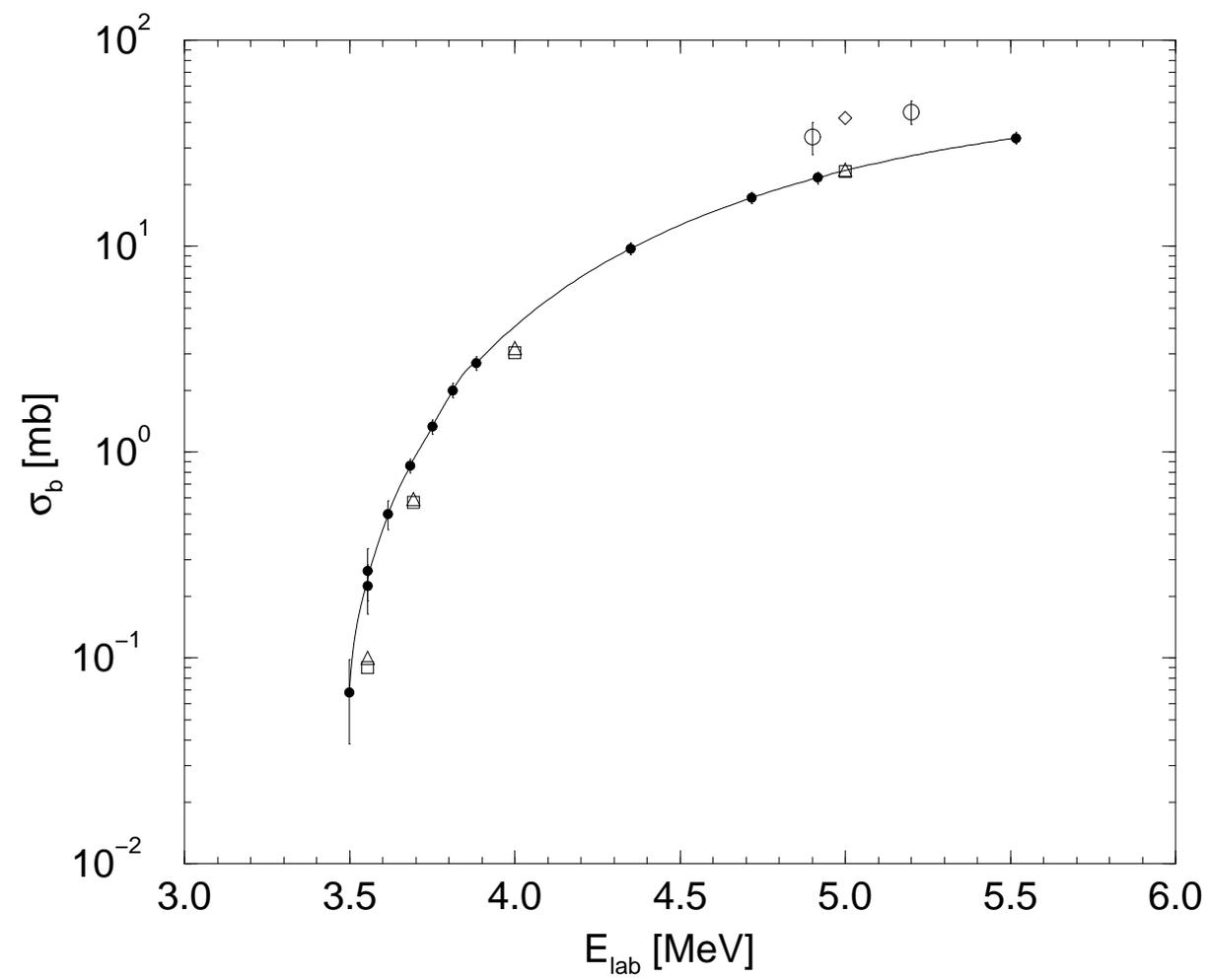
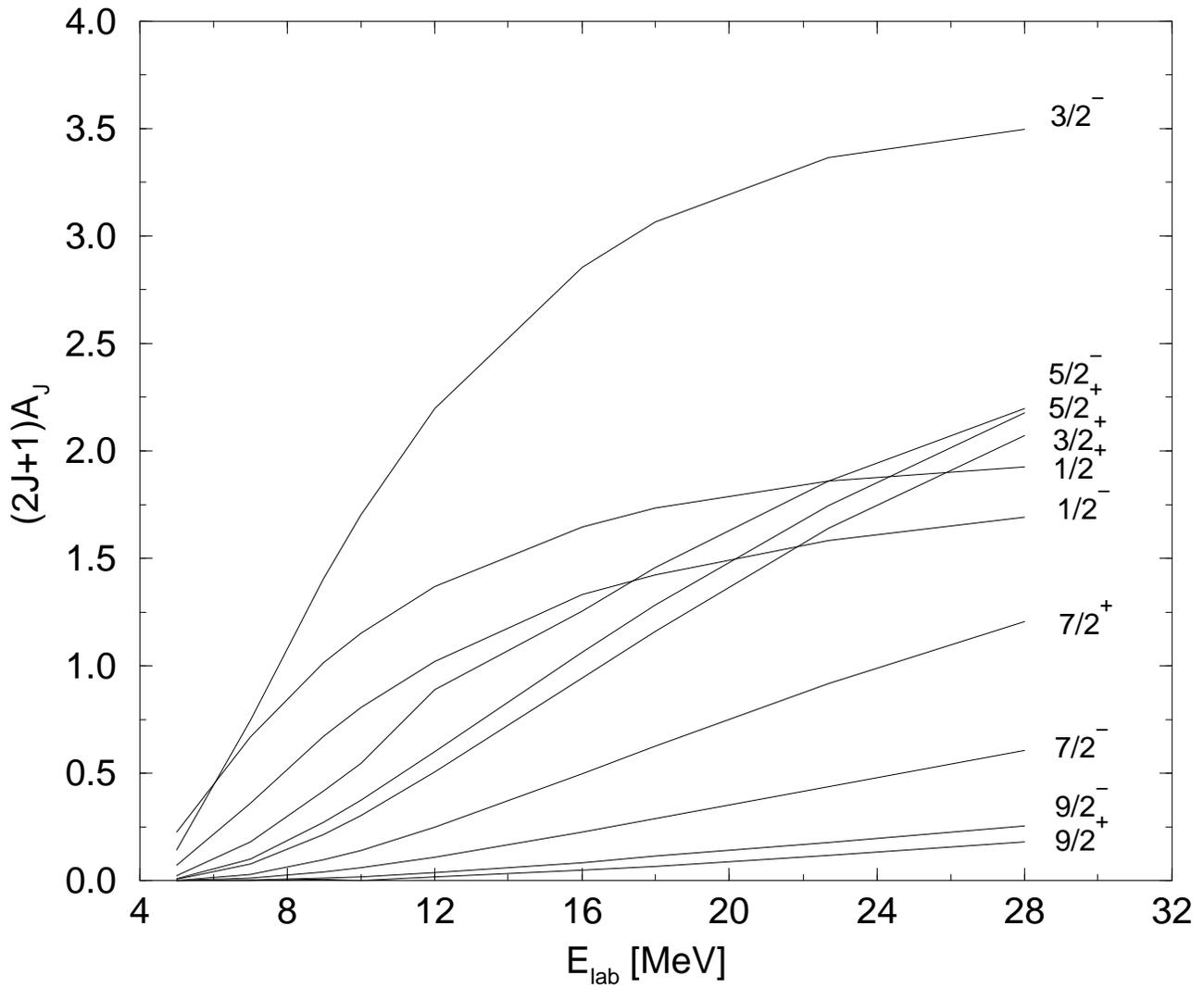
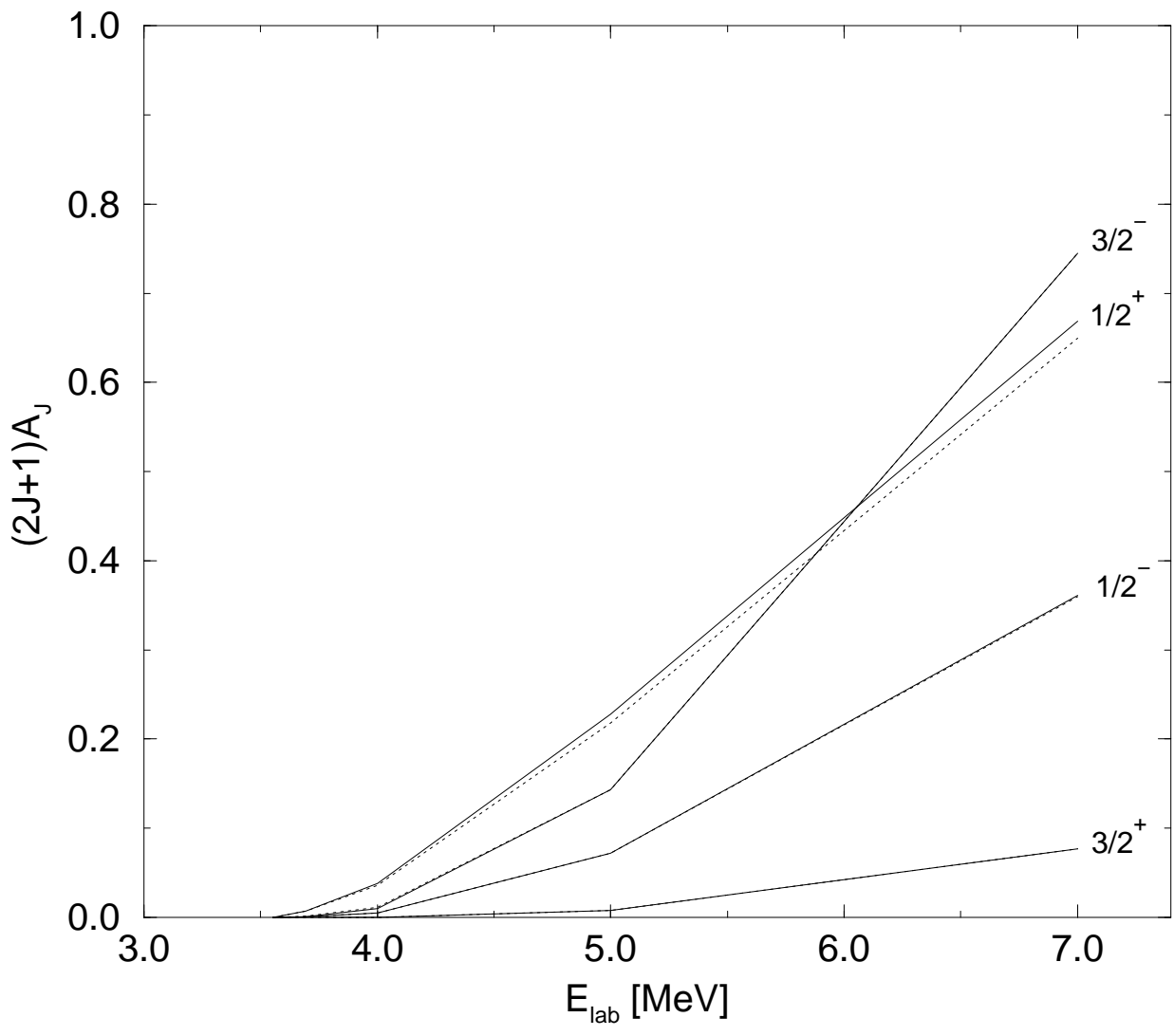
\end{document}